\documentclass[11pt]{article}
\usepackage[letterpaper,textwidth=160mm,hcentering,vmargin=15mm]{geometry} %
\usepackage[T1]{fontenc}
\usepackage[utf8]{inputenc}
\usepackage{textcomp}
\usepackage{mathptmx}
\usepackage{amsmath}
\usepackage{graphicx}
\usepackage{xcolor}
\usepackage{booktabs,tabularx,multirow}
\usepackage{siunitx}
\usepackage[numbers,super,sort&compress]{natbib}
\usepackage{setspace}
\usepackage{caption}
\usepackage[colorlinks=true,linkcolor=black,citecolor=black,urlcolor=blue!50!black]{hyperref}

\newcommand{\sigp}{\ensuremath{\sigma^{+}}}
\newcommand{\sigm}{\ensuremath{\sigma^{-}}}
\newcommand{\fgp}{\ensuremath{f_{\mathrm{g}}^{+}}}
\newcommand{\fgm}{\ensuremath{f_{\mathrm{g}}^{-}}}
\newcommand{\fep}{\ensuremath{f_{\mathrm{e}}^{+}}}
\newcommand{\fem}{\ensuremath{f_{\mathrm{e}}^{-}}}

\newcommand{\mainfig}[3]{%
  \IfFileExists{figures/#1.pdf}{\makebox[\textwidth][c]{\includegraphics[scale=1]{figures/#1.pdf}}}{%
    \fbox{\parbox[c][0.42\textheight][c]{\dimexpr\textwidth-2\fboxsep-2\fboxrule}{\centering\small\textbf{#1} (not yet produced)\\[4pt]#3}}}}

\DeclareCaptionLabelSeparator{bar}{ $|$ }
\newcommand{\snNV}{7} %
\newcommand{\snES}{6} %
\newcommand{\snRatios}{2} %
\newcommand{\snPower}{2} %
\newcommand{\snField}{1} %
\newcommand{\snOblique}{4} %
\newcommand{\snGround}{5} %
\newcommand{\snSim}{3} %
\newcommand{\snLex}{3} %
\newcommand{\sfFourier}{3} %
\newcommand{\sfDrivech}{2} %
\newcommand{\sfGround}{4} %
\newcommand{\sfSwap}{1} %
\newcommand{\sfBand}{5} %
\newcommand{\sfCal}{6} %
\newcommand{\stSim}{1} %
\newcommand{\supfig}[1]{Supplementary Fig.~#1}

\newcommand{\supnote}[1]{Supplementary Note~#1}
\newcommand{\suptab}[1]{Supplementary Table~#1}
\newcommand{\um}{\textmu m}

\begin{document}
\setlength{\emergencystretch}{1.5em}
\doublespacing
\thispagestyle{empty}
\begin{center}
{\LARGE\bfseries Chirality-resolved imaging of spin waves with a quantum spin\par}
\vspace{0.8em}
{\large Enbo Zhang$^{1}$, Tiancheng Zheng$^{1}$, Bo-Yuan Yang$^{1}$, Shreyes Nallan$^{1}$, Chengchao Xu$^{1}$, Mark Ku$^{2}$ and Jian-Gang Zhu$^{1,*}$\par}
\vspace{0.3em}
{\small\singlespacing
$^{1}$Data Storage Systems Center, Department of Electrical and Computer Engineering, Carnegie Mellon University, Pittsburgh, PA, USA\par
$^{2}$University of Delaware, Newark, DE, USA\par
$^{*}$e-mail: jzhu@ece.cmu.edu\par
}
\end{center}
\vspace{0.5em}

\section*{Abstract}
Spin waves carry their propagation direction in the handedness of their circularly polarized stray field, a property inaccessible to scalar magnetometers. In the Damon--Eshbach surface wave of an in-plane magnetized film, the dipolar field above the film rotates with a handedness fixed by the direction of travel. The nitrogen-vacancy (NV) centre in diamond is an intrinsic polarization analyser: a right-handed field about the NV axis drives one spin transition, and a left-handed field drives the other, in both the ground and the excited state. Here, we exploit the two excited-state transitions, which lie inside the spin-wave band of yttrium iron garnet (YIG), as a pair of chirality channels and image Damon--Eshbach waves in both directions. Because reflection from an etched boundary reverses the handedness, the incident and reflected waves appear as two lines in one spectrum and are imaged separately. In the ground state, the same rule applies, and the selected field drives Rabi oscillations, a route to chirality-selective control of the NV spin by magnons. Reading the two channels gives the direction of a spin wave at one point, without a phase reference. This advances chirality-resolved quantum sensing and provides insight into chiral spin-wave dynamics.

\bigskip

\section*{Introduction}
Spin waves, the collective precession of the magnetization whose quanta are magnons, carry information in magnonic circuits in both directions and through nonreciprocal elements\cite{stancil2009spinwaves,chumak2015natphys}. The Damon--Eshbach surface wave of an in-plane magnetized film carries its direction in the polarization of its stray field\cite{damon1961jpcs,kalinikos1986jphysc,kostylev2013jap,yu2019prl}. Its magnetization precesses in the plane of the wavevector and the film normal, and the dipolar field above the film rotates on a circle in the same plane. The direction of this rotation about the bias field, which we call the handedness or chirality of the field, is determined by the sign of the wavevector and by the side of the film. A left-going wave produces a right-handed field above the film, and a right-going wave a left-handed one (as in Fig.~\ref{fig:1}a), although the magnetization precesses in the same direction in both cases\cite{yu2019prl,yu2023physrep}. The same product of signs sets the amplitude factor $1\pm C$, strong on one side of the source and weak on the other, which is the Damon--Eshbach nonreciprocity\cite{kostylev2013jap}.

Conventional metrology does not read the field handedness. Brillouin light scattering, time-resolved Kerr and X-ray microscopy, and inductive spectroscopy record amplitude or phase and infer direction from a phase gradient, a time sequence or a Stokes asymmetry\cite{sebastian2015frontphys,mansfeld2012prl,wintz2016nnano,vlaminck2008science}. A sensor that reads the handedness of a dynamic field directly, with nanoscale spatial resolution and as a channel rather than as a side effect on the signal strength, has not been demonstrated for propagating spin waves.

The nitrogen-vacancy (NV) centre in diamond is such a sensor. Its spin transitions obey a circular selection rule: a microwave field right-handed about the NV axis (\sigp) drives $m_s=0\leftrightarrow+1$, whereas a left-handed field (\sigm) drives $m_s=0\leftrightarrow-1$\cite{alegre2007prb,london2014pra,mrozek2015apl}. The NV centre is therefore an intrinsic polarization analyser, a chirality analyser for the field. The selection rule holds in both triplet manifolds of the NV centre: the ground state with zero-field splitting $D=2.87$\,GHz (transitions \fgp{} and \fgm) and the orbital excited state with $D=1.42$\,GHz (\fep{} and \fem)\cite{fuchs2008prl,batalov2009prl,neumann2009njp,rogers2009njp}. A bias field splits each pair, so that handedness appears as frequency. For an NV axis with a positive projection on the bias field, \sigp{} is the upper line. Read out by optically detected magnetic resonance (ODMR), the four lines form two chirality channels. The line that responds reveals the handedness of the local field and hence, for a known bias-field direction, the propagation direction, at a single point.

NV magnetometry, with NV ensembles and with scanning single spins of nanoscale resolution at room temperature\cite{degen2017rmp,casola2018natrevmater,barry2020rmp}, has detected ferromagnetic resonance (FMR), the magnon chemical potential and propagating spin waves in the low-damping insulator yttrium iron garnet (YIG)\cite{wolfe2014prb,vandersar2015natcommun,du2017science,leewong2020nanolett,bertelli2020sciadv,simon2021nanolett,zhou2021pnas,borst2023science,cheng2026arxiv,vree2026arxiv} and permalloy\cite{manasvalero2026natcommun}.

So far, the NV centre has been read as an amplitude sensor through one ground-state line. The handedness, either of the NV transition or of the stripline excitation, has entered only as the reason why that line is visible on one side of the source\cite{bertelli2020sciadv,luthi2025rsi,simon2021nanolett,prananto2021prapplied}, or as a reference under field reversal\cite{ogawa2025prb}. Polarization-selective NV driving has analysed antenna fields\cite{london2014pra,appel2015njp,wang2015natcommun} and vortex dynamics\cite{rable2026arxiv}. However, it has not imaged the two chiralities of propagating waves. In YIG, the Damon--Eshbach (dipolar) band extends from the FMR to its dipolar edge, $\gamma(B+\mu_0M_\mathrm{s}/2)$. Because $D=2.87$\,GHz exceeds $\gamma\mu_0M_\mathrm{s}/2\approx2.46$\,GHz, \fgp{} lies above the dipolar band at every field used, and \fgm{} near its upper edge, where only short waves of weak amplitude exist (\supnote{\snES}). Ground-state imaging therefore reaches one chirality, on the strong side. Below about 9\,mT, both excited-state lines lie inside the band, at small wavevectors that the stripline launches efficiently. They have been observed in continuous-wave (CW) ODMR spectra above YIG\cite{bertelli2020sciadv,luthi2025rsi}, but their role has remained unclear. To our knowledge, this work is the first to use the two excited-state transitions as a pair of chirality channels, imaging the polarized stray field in both propagation directions, including the weak $(1-C)$ side in its own channel.

Here, we demonstrate chirality-resolved quantum sensing by scanning single-NV magnetometry above a YIG thin film (Fig.~\ref{fig:1}). On the left side of the stripline, \fep{} (\sigp) responds, whereas on the right side \fem{} (\sigm) does. The spectra follow both lines through the band. We find that reversing the bias field exchanges the channels, that the contrast of each channel decays with height as set by its own wavevector, and that the stronger signal lies on the $(1+C)$ side of the nonreciprocity (Fig.~\ref{fig:2}). Near an etched pit, the incident and reflected waves appear as two lines 285\,MHz apart in one spectrum, and maps in the \sigp{} and \sigm{} channels image them separately, in agreement with a micromagnetic model (Fig.~\ref{fig:3}). Curved trenches do the same for obliquely reflected waves (Fig.~\ref{fig:4}). The ground state follows the same rule in the \fgm{} channel, in which the selected field drives Rabi oscillations of the NV spin (\supfig{\sfGround}).

\section*{Results}
\subsection*{The NV centre as a chirality analyser for the spin-wave stray field}
\begin{figure}[!ht]
\centering
\mainfig{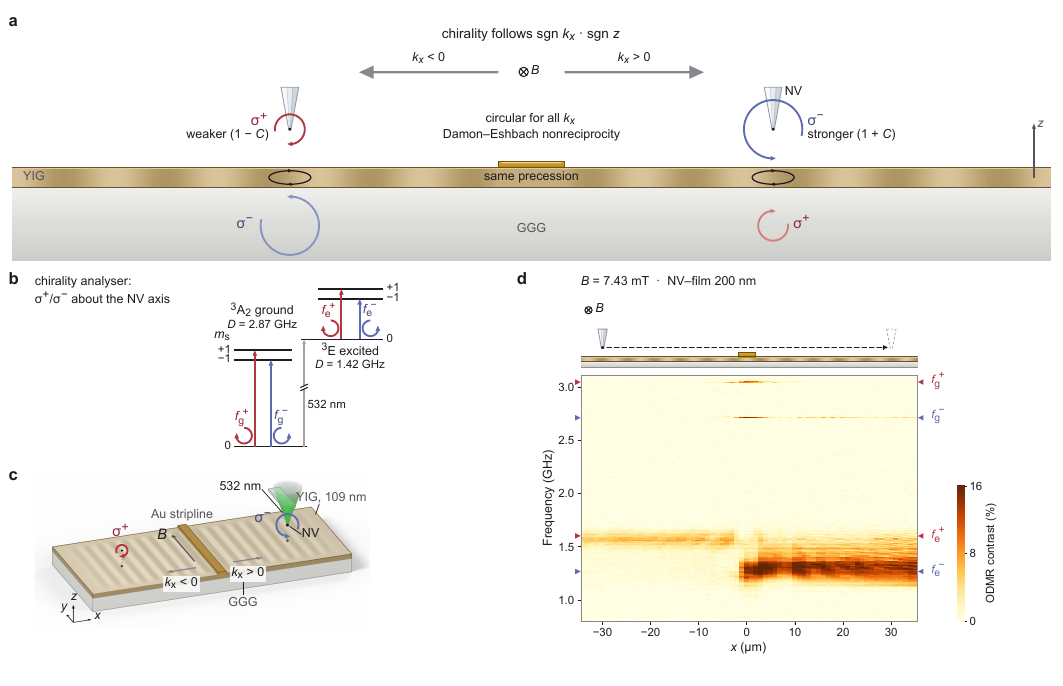}{0.96\textwidth}{Geometry, level scheme, apparatus and lateral ODMR line scan}
\caption{\textbf{The NV centre as a chirality analyser for the Damon--Eshbach stray field.} \textbf{a}, Cross-section viewed along $+B$ ($\otimes B$; $x$ to scale, heights not). The stripline launches Damon--Eshbach surface spin waves toward $-x$ (`$k_x<0$') and $+x$ (`$k_x>0$'), and the magnetization precesses in the same direction on both sides (`same precession', ellipses). Outside the film, the stray field of each wave is circularly polarized for every wavevector (`circular for all $k_x$', equation~(\ref{eq:2})), and its chirality, the direction of rotation about $B$, `follows $\mathrm{sgn}\,k_x\cdot\mathrm{sgn}\,z$': \sigp{} for $k_x<0$ and \sigm{} for $k_x>0$ above the film, and the opposite below the film. Ring radii at equal precession amplitude, `weaker ($1-C$)' and `stronger ($1+C$)' with $C=0.32$, show the `Damon--Eshbach nonreciprocity'.
\textbf{b}, `Chirality analyser: \sigp/\sigm{} about the NV axis'. Spin levels $m_s$ of the $^3\mathrm{A}_2$ ground state ($D=2.87$\,GHz) and the $^3\mathrm{E}$ excited state ($D=1.42$\,GHz), populated by the `532\,nm' laser. A field rotating right-handed about the NV axis, taken with a positive projection on $B$, drives $m_s=0\to+1$, the upper line of each pair, \sigp{} (\fgp{}, \fep{}). Conversely, a left-handed field drives $m_s=0\to-1$, the lower line, \sigm{} (\fgm{}, \fem{}). The line that responds reveals the chirality of the local field. Icons viewed along $+B$.
\textbf{c}, Apparatus (schematic, not to scale). A diamond probe with the NV centre at its apex scans above the YIG film on a GGG substrate, and $B$ lies in-plane along the `Au stripline'. The two rings represent the stray-field chirality at the two probe positions.
\textbf{d}, ODMR contrast versus frequency and position $x$ across the stripline at $B=7.43$\,mT, NV--film 200\,nm. Above the stripline, the Oersted field drives both ground-state lines. Thirty micrometres away, one channel remains per side: \fep{} (\sigp{}) on the left and \fem{} (\sigm{}) on the right, as the chirality rule of \textbf{a} and \textbf{b} predicts.}
\label{fig:1}
\end{figure}

A stripline patterned on a 109\,nm YIG film launches Damon--Eshbach surface spin waves to both sides, with the bias field $B$ along the stripline ($y$) and propagation along $\pm x$ (Fig.~\ref{fig:1}a--c).

For a film magnetized along $+y$, a surface wave of wavevector $k_x$ has the dynamic magnetization
\begin{equation}
\frac{\mathbf{m}(x,t)}{M_\mathrm{s}a}=\cos(k_xx-\omega t)\,\hat{\mathbf{e}}_x+C\sin(k_xx-\omega t)\,\hat{\mathbf{e}}_z,
\label{eq:1}
\end{equation}
with in-plane amplitude $a$ (relative to the saturation magnetization $M_\mathrm{s}$), angular frequency $\omega$ and ellipticity $C$. The ellipticity arises from the demagnetizing field normal to the film plane, which compresses the circular spin precession into an ellipse. The spin precession is right-handed about $+y$ for both propagation directions ($C>0$, Fig.~\ref{fig:1}a). Outside the film ($|z|>d/2$, $z$ from the mid-plane, $d$ the thickness), the dipolar field of this precession is, for $|k_x|d\ll1$,
\begin{equation}
\mathbf{B}=-\frac{\mu_0M_\mathrm{s}a\,d\,|k_x|}{2}\,e^{-|k_x||z|}\bigl(1+C\,\mathrm{sgn}\,k_x\,\mathrm{sgn}\,z\bigr)\Bigl[\cos(k_xx-\omega t)\,\hat{\mathbf{e}}_x-\mathrm{sgn}\,k_x\,\mathrm{sgn}\,z\,\sin(k_xx-\omega t)\,\hat{\mathbf{e}}_z\Bigr]
\label{eq:2}
\end{equation}
(\supnote{\snField}). Outside the film, the field derives from a magnetostatic potential decaying as $e^{-|k_x||z|}$, so $B_z$ and $B_x$ are equal in magnitude and in quadrature. Therefore, the stray field is circular in the $x$--$z$ plane for every $k_x$ and any ellipticity, with handedness fixed by $\mathrm{sgn}\,k_x\cdot\mathrm{sgn}\,z$. Above the film, for $B$ along $+y$, waves toward $-x$ produce a right-handed (\sigp) field and waves toward $+x$ a left-handed (\sigm) field, whereas below the film the handedness is reversed (Fig.~\ref{fig:1}a). The ellipticity appears only in the amplitude factor $1\pm C$, the Damon--Eshbach nonreciprocity, which favours the $+x$ wave above the film ($C=0.26$--$0.33$ here, \supnote{\snField}).

Figure~\ref{fig:1}b shows how the NV centre analyses this handedness. A microwave field rotating right-handed about the NV axis drives the $m_s=0\to+1$ transition, a left-handed field drives $m_s=0\to-1$, and a linearly polarized field drives both transitions\cite{doherty2013physrep,alegre2007prb,london2014pra}. The four lines, \fgp{} and \fgm{} of the ground state and \fep{} and \fem{} of the orbital excited state, which is populated while the laser is on (\supnote{\snES}), form the two chirality channels: \sigp{}, the upper line of each pair, and \sigm{}, the lower line, labelled by their handedness about $B$. In our set-up, the NV axis is tilted from $B$ (Methods), so it sees the circular stray field as an ellipse with a small component of the opposite handedness. This crosstalk contributes to the weak opposite line and produces, in the reflection maps below, interference fringes (\supnote{\snField}).

A lateral ODMR scan across the stripline at 7.43\,mT, 200\,nm above the film, shows the analyser at work (Fig.~\ref{fig:1}d). The contrast is the fractional drop of the NV photoluminescence under resonant microwave drive. On the left of the stripline, the broad \sigp{} line \fep{} at 1.56\,GHz dominates. Conversely, on the right, the \sigm{} line \fem{} at 1.3\,GHz and the narrow ground-state line \fgm{} at 2.71\,GHz appear. The handedness reversal across the stripline is the sign change of $k_x$ in equation~(\ref{eq:2}). Directly above the stripline, its linearly polarized Oersted field drives both ground-state lines. At any point above the film, the line that responds thus reveals whether the spin wave travels toward $+x$ or $-x$, without knowledge of where it was launched.

\subsection*{Two chirality channels left and right of the stripline, swapped by field reversal}
\begin{figure}[p]
\centering
\mainfig{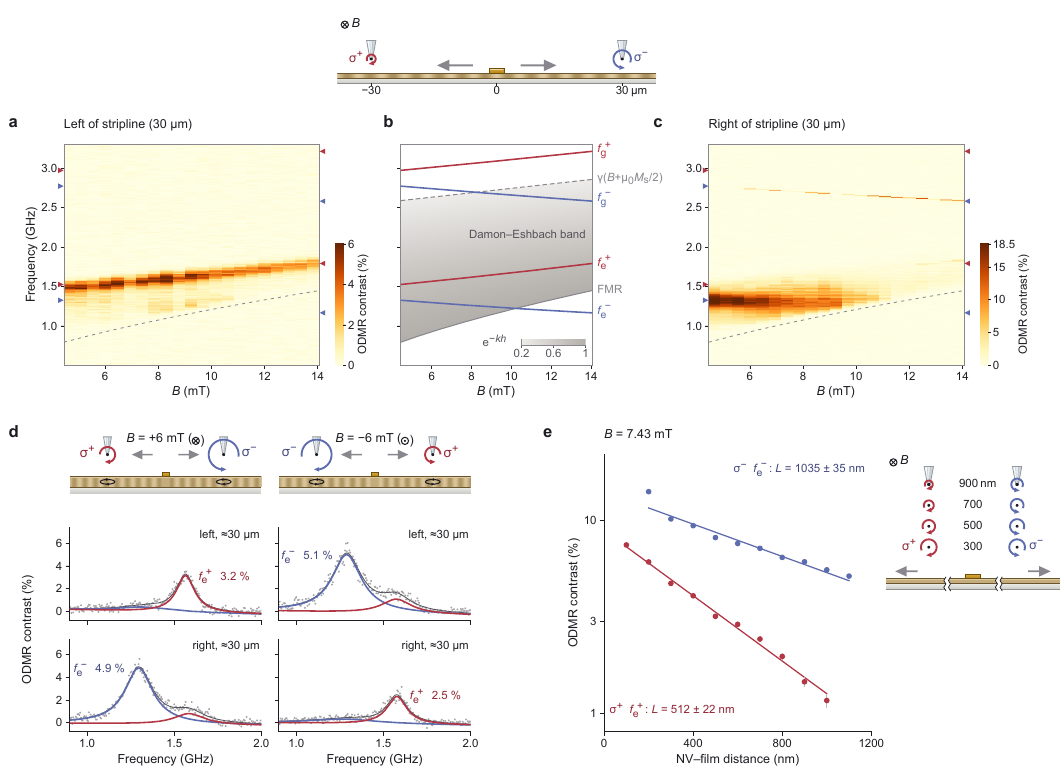}{\textwidth}{ODMR spectra versus field on both sides of the stripline, the calculated band, the field reversal and the height decay}
\caption{\textbf{Two chirality channels left and right of the stripline, swapped by field reversal.} \textbf{a},\textbf{c}, ODMR contrast versus frequency and field, left of the stripline (\textbf{a}) and right of it (\textbf{c}), 30\,\um{} away; NV--film 200\,nm. Top, sketch of the two positions and of the stray-field chirality at each ($\otimes B$). On the left, the \sigp{} channel \fep{} responds; on the right, the \sigm{} channel \fem{} dominates (with a faint \fgm{}). Triangles, NV transitions. Dashed line, FMR.
\textbf{b}, Calculated NV transitions. The shaded Damon--Eshbach band extends from the FMR (solid) to $\gamma(B+\mu_0M_\mathrm{s}/2)$ (dashed); the shading gives the stray-field factor $e^{-kh}$ at the 200\,nm height of \textbf{a} and \textbf{c} (key). At low fields, both excited-state transitions lie inside the band, while \fgp{} lies above it and \fgm{} near its top.
\textbf{d}, ODMR spectra on opposite sides of the stripline under field reversal, at $B=+6$\,mT ($\otimes$) and $-6$\,mT ($\odot$), all four spectra at the same drive power. Lines, two-Lorentzian fits (grey, total), named by frequency order and labelled with the larger contrast: `\fep{} 3.2\%' left and `\fem{} 4.9\%' right at $+6$\,mT; `\fem{} 5.1\%' left and `\fep{} 2.5\%' right at $-6$\,mT. Reversing $B$ reverses the chirality counted about $B$ but not the rotation of each stray field in the laboratory (sketches), so the channels swap sides.
\textbf{e}, ODMR contrast of \fep{} (left of the stripline) and \fem{} (right of it) versus NV--film distance at $B=7.43$\,mT, 30\,\um{} from the stripline. Lines, exponential fits over all recorded heights (100--1,000\,nm for \fep{}, 200--1,100\,nm for \fem{}), which give `\sigm{} \fem{}: $L=1035\pm35$\,nm' and `\sigp{} \fep{}: $L=512\pm22$\,nm'. Inset, sketch of the \sigp{} and \sigm{} stray fields shrinking with height (not to scale). Error bars, standard error of the fit.}
\label{fig:2}
\end{figure}

We recorded ODMR spectra 30\,\um{} left and right of the stripline between 5 and 15\,mT nominal (Fig.~\ref{fig:2}a,c). On the left, the \sigp{} line \fep{} predominates and follows its Zeeman shift. On the right, the \sigm{} lines \fem{} and, from 6\,mT, \fgm{} appear. Below 8.2\,mT, the latter lies just above the dipolar edge and is driven by dipole--exchange waves ($k\approx8$--9\,rad\,\um$^{-1}$, equation~(\ref{eq:3}) in Methods). Each line carries contrast only above the FMR, and the \fem{} signal vanishes once that line crosses below it.

Figure~\ref{fig:2}b places the calculated NV transitions on the Damon--Eshbach band of the film. Both excited-state lines lie inside the band, at small wavevectors that the stripline launches efficiently, whereas \fgm{} lies near the dipolar edge and \fgp{} above it. Higher in the band, the waves are shorter, are launched less efficiently and decay faster with height (grey level in Fig.~\ref{fig:2}b; \supnote{\snES}). The contrast sits on the NV lines, follows them in field and vanishes below the FMR. It therefore arises from resonant, polarization-selective driving of the transitions by the circularly polarized stray field, not from relaxation of the ground-state spin by the magnon bath\cite{wolfe2014prb,du2017science,mccullian2020natcommun}. This noise is itself chiral\cite{rustagi2020prb}, but a relaxation rate does not single out one coherent wave and its direction.

At 5--7\,mT, the ratio of the \fem{} to the \fep{} contrast inverts between the two sides of the stripline by a factor of 90 or more (\supnote{\snRatios}). Because this double ratio compares each line with itself on the two sides, the part of the launch efficiency common to both sides, propagation loss and instrument response cancel, whereas a sensor measuring field strength alone would give only $\approx$0.8, from the factor $1\pm C$. As equation~(\ref{eq:2}) requires, the stronger channel, \sigm{} on the right, is the $+x$ wave favoured by $1+C$.

To test this assignment, we reversed the bias field at $\pm6$\,mT nominal (Fig.~\ref{fig:2}d). The rotation of the stray field in the laboratory, set by $k_x$ and $z$ alone in equation~(\ref{eq:2}), does not change, but the reversed field exchanges the frequency order of the NV transitions. About the reversed field, each wave therefore has the opposite chirality and drives the other line. On the right of the stripline, the response indeed moves from \fem{} (\sigm) to \fep{} (\sigp) when the field is reversed, and the left side shows the opposite exchange (Methods). Because the same positions are compared at nearly the same $|B|$, this test excludes asymmetries of sample, stripline or probe that do not change with the field direction.

The height dependence tests the evanescent form of equation~(\ref{eq:2}) (Fig.~\ref{fig:2}e). Over all recorded heights (100--1,000\,nm for \sigp{}, 200--1,100\,nm for \sigm{}), the \sigp{} contrast of \fep{} falls exponentially with a decay length $L=512\pm22$\,nm, and the \sigm{} contrast of \fem{} with $L=1,035\pm35$\,nm (\supnote{\snField}). The ratio of the two decay lengths, $2.02\pm0.11$, is close to the inverse ratio of the wavevectors at the line centres, 2.26 (1.14 and 0.504\,rad\,\um$^{-1}$). Each decay length is shorter than $1/k$ because the contrast grows with the drive field to a power $n$ (\supnote{\snPower}), so that $L=1/(nk)$. Each channel thus decays with the wavevector of its own wave, which is consistent with the stray field of a propagating wave.

\subsection*{Polarization-selective imaging of incident and reflected spin waves}
Reflection tests the chirality rule with the NV, the bias field and the direction of precession all held fixed. A reflected Damon--Eshbach wave keeps the precession direction of equation~(\ref{eq:1}), which the bias field sets, but the sign of $k_x$ changes. By equation~(\ref{eq:2}), the handedness of its stray field above the film therefore reverses. On the side where the launched wave travels toward $-x$, its stray field drives the \sigp{} channel, whereas the field of its reflection drives the \sigm{} channel. Although the two waves overlap in the film, their counter-rotating stray fields should drive two different transitions of an NV centre at one point.

\begin{figure}[p]
\centering
\mainfig{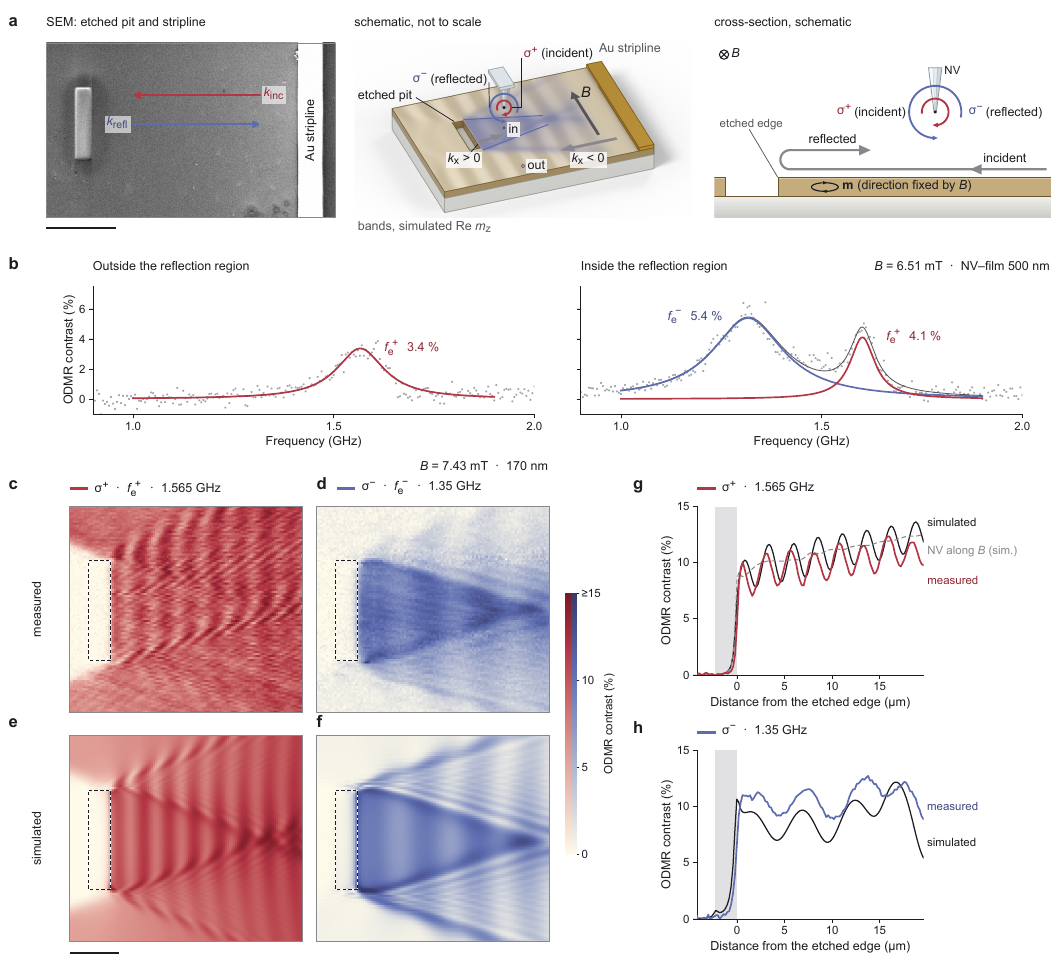}{\textwidth}{SEM, schematic and cross-section of the etched rectangle, spectra outside and inside the reflection region, measured and simulated maps of both channels, and axial profiles}
\caption{\textbf{Reflection reverses the chirality: incident and reflected waves occupy separate channels at one point.} \textbf{a}, Left, SEM image of the etched pit and stripline; middle, schematic of the spin-wave reflection and detection (not to scale); right, cross-section schematic: both waves share the precession direction, which is fixed by the bias field, so reversing $k_x$ reverses the chirality above the film.
\textbf{b}, ODMR spectra measured outside and inside the reflection region, at the same distance from the stripline and at $B=6.51$\,mT. Outside, only the \sigp{} channel appears (\fep{}); inside, the \sigm{} channel of the reflected wave appears as well (\fem{} and \fep{}), about 285\,MHz apart.
\textbf{c},\textbf{d}, Polarization-selective imaging at $B=7.43$\,mT: the \sigp{} channel images the incident wave (\textbf{c}), and the \sigm{} channel images the reflected wave, a converging beam (\textbf{d}).
\textbf{e},\textbf{f}, Simulated maps.
\textbf{g},\textbf{h}, ODMR contrast along the pit axis in \textbf{c} and \textbf{d}, versus distance from the etched edge; grey band, the etched pit. The fringes in \textbf{g} arise from the reflected wave entering the \sigp{} channel through the tilt of the NV axis. Scale bars, 10\,\um{} (\textbf{a}) and 5\,\um{} (\textbf{c}--\textbf{f}).}
\label{fig:3}
\end{figure}

To create a controlled reflection geometry, we etched a rectangular pit through the film on the \sigp{} side, parallel to the stripline (Fig.~\ref{fig:3}a). Its near edge lies 30\,\um{} from the stripline (dimensions in Methods). A cut through the film is needed, because Damon--Eshbach waves pass shallow surface defects almost without back-scattering\cite{mohseni2019prl}. In thick films, reflection at a film end proceeds through thickness-quantized modes\cite{vasyuchka2026prl}, which lie above the band in our 109\,nm film. In the simulation, an edge that cuts through the film at $kd\approx0.1$ returns the wave into the counter-propagating branch (\supnote{\snSim}).

Spectra at two positions at the same distance from the stripline, outside and inside the reflection region, confirm the chirality reversal (Fig.~\ref{fig:3}b). Outside, only \fep{} appears, with 3.4\% contrast, as for the unperturbed incident wave. Inside, a \sigm{} line (\fem{}, 5.4\%) appears 285\,MHz below the \sigp{} line (\fep{}, 4.1\%). The reflected wave therefore appears at the same point, in the opposite channel of the same spectrum.

Mapping the contrast at each transition images the two chiralities separately (Fig.~\ref{fig:3}c,d). Because of the splitting, the two maps sample different wavevectors, $k=1.14$\,rad\,\um$^{-1}$ at 1.565\,GHz and 0.60\,rad\,\um$^{-1}$ at 1.35\,GHz (equation~(\ref{eq:3}) in Methods). The \sigp{} map at \fep{} shows the incident wave. Its contrast is uniform away from the pit and carries standing-wave fringes parallel to the edge near it, as in magneto-optical images of spin-wave reflection and refraction at edges and steps\cite{stigloher2016prl,stigloher2018prl}. The \sigm{} map at \fem{} shows predominantly the reflected wave, as a converging beam that leaves the pit edge. This beam is the near-field diffraction pattern of a reflector about one wavelength long ($\lambda=10.5$\,\um{} at 1.35\,GHz). It carries about seven times the contrast measured beside it, where only the incident wave passes (\supnote{\snSim}).

The channel selects by chirality, not by strength. To test this, we simulated the full geometry with a two-dimensional micromagnetic model (Fig.~\ref{fig:3}e--h; Methods and \supnote{\snSim}), which converts the stray field at the NV into contrast by the chirality rule of Fig.~\ref{fig:1}b. In the simulated drive field, before any NV response, the reflected wave carries the stronger stray field inside the beam at both map frequencies, by 3--5\,dB (\supfig{\sfDrivech}). It travels toward $+x$, on the $(1+C)$ side of the nonreciprocity. Nevertheless, the \sigp{} map shows the weaker, incident wave. The simulation reproduces the fringed incident wave in \sigp{} and the converging beam in \sigm{} (Fig.~\ref{fig:3}e,f). The contrast is of the same size: the beam-to-background ratio of the \sigm{} map is $7.4\pm0.7$ measured and 4.1--6.2 simulated. Swapping the channel assignment, as the opposite rule would require, removes the beam from the \sigm{} map (\supfig{\sfSwap}). This ratio then falls to 1.2, and the simulated fringe visibility of the \sigp{} map, discussed next, falls to 0.09, against 0.17 measured (\supnote{\snSim}).

The fringes in the \sigp{} map would be absent for a perfect analyser, because the reflected wave is left-handed and should not enter the \sigp{} channel. They arise from crosstalk through the tilt of the NV axis. About a tilted axis, the circularly polarized $x$--$z$ stray field of the reflected wave contains a small component of the opposite handedness (\supnote{\snField}). This component drives the \sigp{} transition together with the incident wave, and the two waves interfere, so the \sigp{} contrast oscillates along the beam with half the wavelength. We quantify the fringes by their visibility, the contrast difference between bright and dark fringes divided by their sum. Over 0.3--6\,\um{} from the edge, the measured visibility is 0.17, within the 0.15--0.18 that the simulation gives for our tilted NV (Fig.~\ref{fig:3}g). For an NV axis along the field, the simulation gives only 0.04--0.07, from the oblique parts of the beam. The fringes therefore come from the tilt, and their size matches it.

The finite length of the reflecting edge scatters part of the incident surface wave into oblique directions, giving the scattered waves a transverse wavevector component $k_y$ along the magnetization. Because the film's dispersion is anisotropic, their wavelength differs from that of the waves along $x$, which tilts the fringes near the corners. A two-dimensional Fourier transform of each map resolves the $k_x$ and $k_y$ components of the waves in its channel. Each spectrum peaks at $2k_0$, twice the wavevector of the channel's own wave, as the half-wavelength period of its fringes requires\cite{simon2022nanolett,berthomier2026arxiv}, and lies on the isofrequency contour displaced by the incident wavevector (\supfig{\sfFourier}, \supnote{\snOblique}).

\subsection*{Curved reflectors and oblique waves}
\begin{figure}[p]
\centering
\mainfig{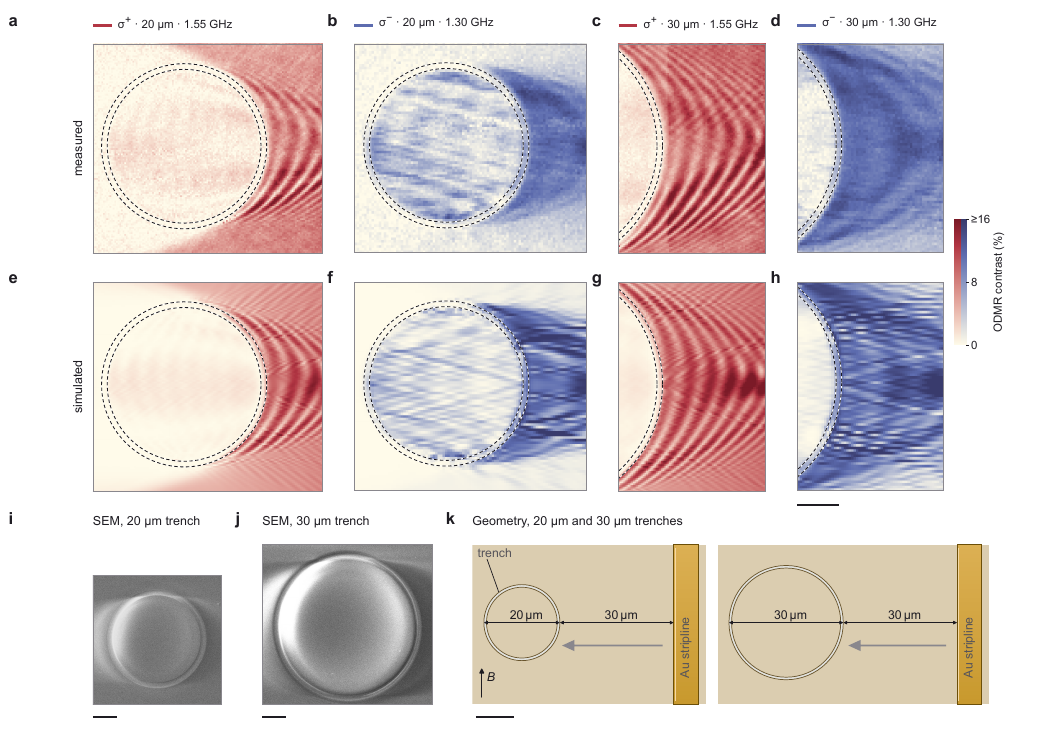}{\textwidth}{Measured and simulated maps of both channels at the 20 and 30 micrometre circular trenches, with SEM panels}
\caption{\textbf{Polarization-selective imaging at circular reflectors: waves along $\pm x$ separate into the two channels, oblique waves appear in both.} \textbf{a}--\textbf{d}, Measured ODMR contrast maps around circular trenches of 20\,\um{} (\textbf{a},\textbf{b}) and 30\,\um{} (\textbf{c},\textbf{d}) outer diameter, each enclosing an intact YIG disk. Dashed circles, the SEM trench outlines used in the model. The crescent on the stripline side in \textbf{b},\textbf{d} is the reflected wave travelling toward $+x$, imaged in the \sigm{} channel alone. Waves along $\pm x$ separate cleanly, one channel each, whereas obliquely reflected waves appear in both (\supnote{\snOblique}).
\textbf{e}--\textbf{h}, Simulated maps of \textbf{a}--\textbf{d}. They reproduce the chevrons, the crescent and their channel assignment, and the presence of the stripes inside the 20\,\um{} disk.
\textbf{i},\textbf{j}, SEM images of the 20\,\um{} (\textbf{i}) and 30\,\um{} (\textbf{j}) trenches. Because YIG is insulating, charging saturates the image along the horizontal beam-scan direction.
\textbf{k}, Schematic of the configuration; $B$ lies in-plane along the stripline. Scale bars, 5\,\um{} (\textbf{a}--\textbf{j}) and 10\,\um{} (\textbf{k}).}
\label{fig:4}
\end{figure}

A curved wall reflects waves into all directions and therefore tests the channels with waves that do not travel along $\pm x$. The stray field of a surface wave rotates about the in-plane axis normal to its wavevector. For waves close to $\pm x$, this axis lies along the bias field, closest to the NV axis, so the two channels are cleanest there. Oblique waves appear in both channels (\supnote{\snOblique}). We etched circular trenches around two intact YIG disks of 20 and 30\,\um{} outer diameter, 30\,\um{} from a second stripline. Unlike the rectangular pit, the curved walls have no sharp corners, at which the demagnetizing field is strong. We mapped them at 1.55\,GHz (\sigp{}, Fig.~\ref{fig:4}a,c) and 1.30\,GHz (\sigm{}, Fig.~\ref{fig:4}b,d), at 6.51\,mT and an NV--film distance of 200\,nm (geometry in Fig.~\ref{fig:4}i--k).

In the \sigp{} maps, the incident wave carries fringes in front of each disk, and V-shaped chevrons of obliquely reflected waves appear around it. The disk interior is dark (1.1\% against about 5\% outside the 20\,\um{} disk). In the \sigm{} maps, the reflected wave forms a crescent on the stripline side of each wall, where it travels toward $+x$. This is the same channel separation as at the pit. Inside the 20\,\um{} disk, the \sigm{} map also shows oblique stripes (3.5\%). Because the trench is narrow (about 0.7\,\um{}), part of the incident wave couples across it through its dipolar field and is reflected inside the disk. The simulation reproduces the chevrons, the crescent and the presence of the stripes (Fig.~\ref{fig:4}e--h).

\subsection*{The ground-state channel and coherent driving}
The ground-state transitions obey the same selection rule, but in YIG only one reaches the Damon--Eshbach band. At 13.85\,mT, the \sigm{} transition \fgm{} (2.59\,GHz) lies inside the band and \fgp{} above it (Methods). We observed \fgm{} 30\,\um{} right of the stripline but not on the left (\supfig{\sfGround}b), as expected for the \sigm{} field of a right-going wave. On its own, this test is weak. The waves are short here, and the nonreciprocity alone makes the left-going wave much weaker above the film, so a chirality-blind sensor would also see little on the left.

A map at \fgm{} beside the pit images the reflected wave, whose \sigm{} field drives this line (\supfig{\sfGround}c). Beside the end of the pit, which the incident wave reaches unobstructed, the map shows no contrast, so the signal comes from the reflected beam (\supnote{\snGround}).

The polarized field of the reflected wave is also a control field. With the probe at a contrast maximum in the beam, we drove \fgm{} with microwave pulses of increasing length. The spin population oscillates with the pulse length at $\Omega/2\pi=1.756\pm0.014$\,MHz (\supfig{\sfGround}d): Rabi oscillations, the signature of coherent driving. Earlier work drove the NV spin with the launched wave\cite{andrich2017npjqi,wang2020npjqi}; here the drive is the chirality-selected field inside the reflected beam.

\section*{Discussion}
Polarization-selective sensing achieves what spatial resolution alone cannot. Amplitude sensing infers the propagation direction from a phase pattern or time sequence, whereas the NV centre reads it from the chirality of the stray field, at one point under CW drive. The double ratio across the stripline, the second line at the pit and the control simulation with swapped channels separate chirality from strength (Figs~\ref{fig:2} and \ref{fig:3}), and the height decay and the field reversal tie each channel to its own wave.

The ratios between the two lines are channel contrast ratios, not degrees of circular polarization, because the contrast is not proportional to the wave amplitude (\supnote{\snPower}), and their double ratio is large only at low fields. Because the two excited-state lines are broad and a few hundred MHz apart, they image waves of different wavelengths. We tested the ground-state rule in one line and one map (\supnote{\snGround}).

Because the selection rule is an intrinsic property of the NV centre, the sensing scheme transfers to any excitation whose magnetic near field rotates in a direction locked to its wavevector\cite{bliokh2014natcommun,yu2023physrep}, from topological magnon edge modes\cite{shindou2013prb,chisnell2015prl} to spin--orbit-torque-driven precession. Channels become cleaner as the NV axis approaches the bias field, the rotation axis of the stray field. Where both ground-state lines fall inside the band, as in films of higher magnetization such as permalloy or CoFeB, the much narrower ground-state pair replaces the excited-state pair.

Reading the stray-field chirality opens direct access to chiral and nonreciprocal magnonics and, more broadly, to spintronics. In a magnonic circuit\cite{chumak2015natphys,mahmoud2020spinwavecomputing}, the two channels resolve forward and backward waves at edges, junctions, defects and reflectors, where amplitude maps alone do not separate them. With each line calibrated, the channel pair can test nonreciprocal elements and resolve the wavelength asymmetry of interfacial Dzyaloshinskii--Moriya interaction\cite{di2015prl}. With its two transitions, the NV centre is a magnetic analogue of the chiral emitter of quantum optics\cite{lodahl2017nature}, used here as an absorber. Proposals to sense the handedness of magnetic noise\cite{rustagi2020prb} and to couple NV qubits through a film via chiral magnons\cite{xue2025prb,dols2026prresearch} rest on this selection rule. Natural next steps would be emission of a spin wave by the NV and low-temperature operation. Polarization-selective quantum sensing establishes the chirality of the stray field as a sensing dimension of its own for nanoscale magnetic imaging.

\section*{Methods}
\subsection*{Sample and reflectors}
Matesy GmbH grew the YIG film by liquid-phase epitaxy on (111) gadolinium gallium garnet (GGG). The supplier specifies a thickness of 109\,nm with at most 1\% variation, an FMR linewidth below 2\,Oe at 6.5\,GHz and an r.m.s.\ roughness of 0.5\,nm. We use $\mu_0M_\mathrm{s}=176$\,mT, as measured on thicker films grown the same way\cite{klingler2015jphysd}, and an exchange length $\lambda_\mathrm{ex}=20$\,nm (\supnote{\snLex}).

To create the reflectors, we milled the film with a xenon plasma focused ion beam at 30\,keV, with a beam current below 0.1\,nA for the trenches, which were milled to a set depth of 1\,\um. We measured their outline with a Helios 5 Hydra UX scanning electron microscope (SEM). The rectangular pit is 10.2 by 2.32\,\um, with its near edge 29.8\,\um{} from the stripline. Beside a second stripline, circular trenches 0.7\,\um{} wide surround two intact disks. Their outer diameters are 20 and 30.1\,\um.

\subsection*{Stripline and microwaves}
The only microwave source is a stripline of 5\,nm Ti and 100\,nm Au, 500\,\um{} long, made by photolithography, sputtering and lift-off. It is 5\,\um{} wide by design. CW microwaves from an Agilent E8257D generator passed through a Mini-Circuits ZASWA-2-50DRA+ switch and an amplifier. Quoted powers are generator settings. An NdFeB magnet, pre-set with a Hall gaussmeter, applied the field $B$ in the plane along the stripline. All measurements were made at room temperature.

\subsection*{Scanning NV microscope}
We scanned a Qnami diamond probe, whose NV centre lies about 10\,nm below the apex, using a tuning-fork atomic force microscope combined with a confocal microscope. A 532\,nm CW laser, below 100\,\textmu W, excited the NV through a $50\times$ objective with a numerical aperture of 0.55. The photoluminescence passed through a Semrock 552\,nm dichroic mirror and a 647\,nm long-pass filter to an Excelitas SPCM-AQRH-16-FC counter, which recorded 90--210\,kcounts\,s$^{-1}$. After contact, we set the NV--film distance, the height of the NV above the film, by a calibrated retraction. The supplier measured $g^{(2)}(0)<0.5$ for the probe, which indicates a single centre.

\subsection*{ODMR measurements and fitting}
We recorded CW ODMR spectra, mostly from 0.5 to 3.505\,GHz in 5\,MHz steps. Each frequency point averaged 60--400 readings of about 21\,ms. To normalize each spectrum, we divided the photoluminescence by a reference at the same point. The contrast of a line is the depth of a Lorentzian on a linear baseline, and the error of each point is the noise away from resonance, 0.14--0.25\%. We fitted both excited-state lines jointly to obtain their channel contrast ratio.

\subsection*{Field and NV-axis calibration}
To calibrate the field and the NV axis, we fitted both ground-state lines on the stripline at 2--10\,mT nominal, with $\gamma/2\pi=28.03$\,GHz\,T$^{-1}$. The fit gives a zero-field splitting of $2870.6\pm0.2$\,MHz and an NV axis tilted by $36.5\pm0.4^\circ$ from the field (\supfig{\sfCal}, \supnote{\snNV}). The calibrated field, $B=0.917(5)\,B_\mathrm{nom}+0.09(1)$\,mT, converts nominal 7, 8 and 15\,mT into 6.51, 7.43 and 13.85\,mT. We label data by the nominal field but calculate with the calibrated one. We calibrated the field-reversal series (Fig.~\ref{fig:2}d), recorded 500\,nm above the film, in the same way (\supnote{\snNV}).

\subsection*{Maps and line scans}
Each map pixel is the contrast at one frequency and height, computed as for the spectra. We recorded the pit maps with $120\times100$ pixels of $200\times213$\,nm at 1.565 and 1.35\,GHz, 7.43\,mT, 0\,dBm and 170\,nm. The trench maps, with 4,800--16,800 pixels, were recorded at 1.55 and 1.30\,GHz, 6.51\,mT and 200\,nm. A pixel took 2--3\,s, and a map 5--9\,h. The fast axis agrees with SEM within 2\%, and we calibrated the slow axis from the roundness of the trenches. The line scan of Fig.~\ref{fig:1}d used steps of 1\,\um.

\subsection*{Image processing}
SEM images were cropped, rotated by multiples of 90$^\circ$ and downsampled, and their brightness and contrast were adjusted linearly. The Au stripline saturates in Fig.~\ref{fig:3}a, and charging saturates the left wall in Fig.~\ref{fig:4}j. ODMR maps, line scans and spectra show the raw contrast without smoothing.

\subsection*{Spin-wave dispersion}
To relate the angular frequency $\omega$ to the wavevector $k$, $2\pi$ divided by the wavelength, we used the lowest mode of the Kalinikos--Slavin theory\cite{kalinikos1986jphysc,kostylev2013jap}:
\begin{equation}
\omega^2=\bigl(\omega_H+\lambda_\mathrm{ex}^2\omega_Mk^2\bigr)\bigl(\omega_H+\lambda_\mathrm{ex}^2\omega_Mk^2+\omega_MF\bigr),
\label{eq:3}
\end{equation}
with $F=1+\omega_MP(1-P)/(\omega_H+\lambda_\mathrm{ex}^2\omega_Mk^2)$, $P=1-(1-e^{-kd})/(kd)$, $\omega_H=\gamma B$ and $\omega_M=\gamma\mu_0M_\mathrm{s}$. This relation holds for waves travelling perpendicular to the magnetization. The precession ellipticity $C$ of equation~(\ref{eq:1}) is $C=\sqrt{(\omega_H+\omega_MP)/(\omega_H+\omega_M(1-P))}$. Kittel's formula\cite{kittel1948physrev} with the same parameters gives the FMR line in Fig.~\ref{fig:2} and \supfig{\sfBand}. At the pit-map frequencies of 1.565 and 1.35\,GHz, equation~(\ref{eq:3}) gives $k=1.14$ and 0.60\,rad\,\um$^{-1}$. For the height decays (Fig.~\ref{fig:2}e), we take $k$ at the line centres, 1.14\,rad\,\um$^{-1}$ (\fep{}, 1.565\,GHz) and 0.504\,rad\,\um$^{-1}$ (\fem{}, 1.308\,GHz; \supnote{\snField}).

\subsection*{Micromagnetic simulation}
We solved the Landau--Lifshitz--Gilbert equation with our two-dimensional Python solver, which treats the film as one uniform layer (\supnote{\snSim}). The solver uses the film parameters above, a damping of $2.5\times10^{-3}$, the calibrated field and the reflector shapes measured by SEM (\suptab{\stSim}). A current sheet in place of the stripline drives both map frequencies. We projected the stray field at the NV height onto the NV axis, whose in-plane part, of magnitude 0.14 for the nominal (100) orientation of the tip, lies along $-x$. A CW line shape 160\,MHz wide then converts the \sigp{} and \sigm{} drive into contrast.

We matched the mean simulated contrast outside the reflector to the measurement with one amplitude factor per map. For the trenches, one factor per channel, shared by both disks, agrees with separate factors within 0.5\,dB. We compared the fringe visibility 0.3--6\,\um{} from the pit edge and the beam contrast against its surroundings, with the swapped channel assignment and an NV axis along $B$ as null tests (\supnote{\snSim}). We located the pit edge to within $\pm0.08$\,\um{} using the photoluminescence and the SEM width.

\subsection*{Ground-state map and Rabi measurement}
We mapped \fgm{} at 2.59\,GHz, 13.85\,mT. For the Rabi measurement, a 2\,\textmu s laser pulse initialized the spin, and a 2.59\,GHz pulse then drove it for 60--1,980\,ns in 20\,ns steps, before a second laser pulse read it out. A SpinCore PulseBlasterUSB card gated the pulses through the switch. The probe was about 200\,nm above the film, at a contrast maximum in the beam.

Fitting a damped cosine gives $\Omega/2\pi=1.756\pm0.014$\,MHz (\supnote{\snGround}). For a circularly polarized drive, $\Omega=\sqrt2\gamma B_\perp$, so the co-rotating field in the NV frame is about 44\,\textmu T. Because detuning raises the fitted frequency, 44\,\textmu T is an upper bound.

\subsection*{Control experiments}
We tested whether the direct field of the stripline, rather than the spin waves, drives the NV 30\,\um{} away. Two observations rule this out. First, away from the stripline each line appears on one side only, in the channel that the propagation direction selects (Figs~\ref{fig:1}d and \ref{fig:2}a,c), whereas the linearly polarized direct field drives both transitions on both sides, as it does above the stripline. Second, the ground-state line \fgp{} gives a quantitative bound. At 30\,\um{} it lies above the Damon--Eshbach band at every field, so any drive by the direct field would appear in it. Its absence from all 42 spectra of the field series limits the direct field at the NV to below 4--22\,\textmu T, where the range reflects the uncertainty in the saturation field of the ground-state line, whereas the excited-state lines would need 0.15--0.46\,mT to reach their observed contrast (\supnote{\snPower}). The dark region behind the pit gives an independent bound, with 0.06\% measured against 1.0\% simulated for an unscreened stripline (\supnote{\snSim}).

Two measurements bound microwave heating of the film and of the NV. From 0 to 10\,dBm, the cut-off edge of the \sigp{} line at 15\,mT limits any change of $M_\mathrm{s}$ to 1.2\%, corresponding to about 6\,K\cite{hansen1974jap}. At the NV, the shift of \fgm{}, $-0.11\pm0.77$\,MHz, limits the heating to 22\,K\cite{acosta2010prl}.

\subsection*{Statistics and reproducibility}
We recorded field scans on both sides of the stripline at several NV--film distances. Similar results were obtained on further films with the same stripline design and with other NV probes.

We recorded the \sigp{} and \sigm{} maps of each etched pit several times, including once at 1.53 and 1.31\,GHz, 5\,dBm; the repeats have a pixel correlation of 0.97--0.98 with the maps shown. In a repeat of the \sigm{} map at 150\,nm, the beam-to-background ratio is $5.8\pm0.5$, compared with $7.4\pm0.7$. We also mapped pits of other dimensions, which separate the incident and reflected waves into the two channels in the same way. We recorded two Rabi traces; one is shown in \supfig{\sfGround}d. We recorded the four field-reversal spectra at the same drive power, and in each spectrum the weaker line, which lies in the tail of the broad stronger line, stays at or below 1.3\%.

\subsection*{Use of AI tools}
Large-language-model agents (Claude, Anthropic) were used under the direction of E.Z. to script the figures and renders (no generative image model was used). The authors take full responsibility for the content.

\section*{Data availability}
The data that support the findings of this study, including the source data for the figures, are available from the corresponding author upon request.

\section*{Code availability}
The analysis and simulation code is available from the corresponding author upon request.

\bibliographystyle{naturemag}
\bibliography{refs,refs_chiral}

\section*{Acknowledgements}
This work was supported by the Data Storage Systems Center (DSSC) at Carnegie Mellon University and its sponsors. We thank J. A. Bain and Q. Li for helpful discussions. We acknowledge the use of the Materials Characterization Facility at Carnegie Mellon University supported by grant MCF-677785, and of the Claire \& John Bertucci Nanotechnology Laboratory at Carnegie Mellon University.

\section*{Author contributions}
J.-G.Z. and E.Z. conceived the project. E.Z., J.-G.Z. and M.K. developed the methodology. E.Z. built the scanning-NV measurements and performed the experiments with T.Z., B.-Y.Y. and S.N. B.-Y.Y. fabricated the stripline, and E.Z. and T.Z. milled the reflectors with a focused ion beam. E.Z., T.Z. and C.X. analysed the data. J.-G.Z. and M.K. supervised the work. J.-G.Z. acquired the funding. E.Z. and J.-G.Z. wrote the manuscript with input from all authors.

\section*{Competing interests}
The authors declare no competing interests.

\clearpage
\section*{Figure legends}

\end{document}